# Comparison of neutron and ion irradiation induced lattice parameter changes in Ni, Cu and MgO single crystals

Xin JIN[1,2], Alexandre BOULLE[2], Jérôme BOURCOIT[1], Aurélien DEBELLE[1]

1. Université Paris-Saclay, CNRS/IN2P3, Institut de Physique des 2 Infinis Irène Joliot-Curie, IJCLab, 91405 Orsay, France
2. IRCer, CNRS UMR 7315, Centre Européen de la Céramique, 12 rue Atlantis, 87068 Limoges Cedex, France

Data available in the literature of lattice parameter changes in Ni, Cu and MgO single crystals irradiated in neutron reactors are tentatively reproduced using *ad hoc* ion irradiation experiments. The nature and energy of the ions were selected so that their weighted average recoil spectrum matches at best that of the neutron flux inside the reactor. The ion fluence was calculated to obtain similar, average displacements per atom (dpa) levels. The lattice parameters of the irradiated samples were determined using X-ray diffraction. We show that, for the metals, it is not possible to reproduce the neutron data, essentially because the dpa levels are low and the irradiated, damaged volume is small. On the contrary, for MgO, a very close lattice parameter change is obtained, providing that the irradiation temperature is significantly increased (1000 K) with respect to that in the reactor (300 K).

The understanding of radiation effects in materials has gained a lot from the study of nuclear materials because, as these later are inherently submitted to various irradiation sources, the study of their behavior under harsh radiative environments was, and remains an important topic to tackle. The major issue with respect to particle irradiation for nuclear materials is neutron irradiation. In order to investigate the effects of this type of irradiations, there are two options. The first one is to use test reactors that reproduce, quite closely, actual conditions experienced in commercial reactors. The second option lies in the use of ion beams delivered by particle accelerators. Pros and cons of these two alternatives have been recently addressed in a review paper [1] and also in [2-3] and it is not in the scope of the current paper to make such a detailed comparison. For decades, numerous works have been carried out to theorize on the simulation of neutron damage by ion irradiation. The reader can refer to [4-5] and references therein to find more information about fundamental and practical concepts. Despite this extensive work, the question of the validity of ion irradiation as a surrogate to neutron irradiation is still debated. As an evidence of this statement, we can mention an ongoing research work of the International Atomic Energy Agency (IAEA) named "Accelerator Simulation and Theoretical Modelling of Radiation Effects - SMoRE-II". One of the objectives of this program is to quantify the degree of agreement between microstructures generated by ion and neutron irradiation. Indeed, there are various discrepancies between the two methods, particularly in the damage characteristics such as its volume, its creation rate, its homogeneity, to mention the most important ones and these differences can lead to significantly different microstructures whereas the ion irradiation method is supposed to be a proxy for the neutron irradiation one. Nevertheless, it has been shown in several cases that the former can allow reproducing phenomena (e.g. swelling, creep, radionuclides production) occurring during the later in fission and fusion reactors [1,6-8]. Therefore, one can expect that the corresponding microstructural features are also identical, which is sometimes the case [9-11], but not always [12-13]. In fact, it is usually impossible for ion irradiation experiments to simultaneously achieve all aspects of microstructural changes occurring under neutron irradiation [2,10]. In conclusion, the answer to the above-mentioned question is not straightforward.

In the current paper, we aim at providing a piece of answer to that very question. For this purpose, we used a simple strategy. We first collected in the literature data of lattice parameter changes (i.e. elastic strain build-up) in neutron-irradiated single crystals. Then, we carried out targeted ion irradiation experiments (see sect. II.2.), followed by X-ray diffraction (XRD) measurements, to determine the corresponding strain build-ups. We show hereafter that even such a simple comparison between strain development kinetics from neutron and ion irradiation experiments is not direct.

## II. Methodology and experimental details

II.1. Experimental details

Ni, Cu and MgO single crystals were purchased to MTI corporation in Richmond, CA, USA. Purity of the metals is > 99.99 % and that of MgO is > 99.95 %. Lattice parameters of pristine samples were found to be those of the pure, perfect materials, i.e. 0.352 nm, 0.362 nm and 0.422 nm for Ni, Cu and MgO, respectively. Samples were mirror-polished with a very weak surface roughness.

Ion irradiation experiments were conducted on the SCALP platform of the IJCLab using either the 190 kV implantor (for Bi irradiations) or the tandem accelerator for I, Cl, Ni and Au irradiations. All series of irradiations were carried out at room temperature (RT). Details of the ion energy, fluences and ranges are provided in each corresponding sub-sections III.

X-ray diffraction measurements were performed on two devices. Details of the two set-ups are provided in [14]. We simply want to mention here that we used high-resolution configurations for all measurements to ensure to detect the smallest peak shifts. These latter essentially depend on the intrinsic peak width and on the probed hkl reflection, but typically a shift corresponding to a lattice strain of 0.025 % can be detected.

II.2. Methodology

In neutron irradiation, during a collision, the energy transferred by the neutron to a target material atom, usually referred to as the Primary Knock on Atom (PKA) or a recoil, can be sufficient (if it exceeds the threshold displacement energy, $E_d$) to displace this atom. If this later has enough energy, it can in turn displace other target atoms, giving rise to a collision cascade and the total number of displaced atoms depends on the PKA energy [15-18]. All the displaced atoms then evolve and a fraction of them produce (clusters of) defects that modify the material properties. Therefore, in order to accurately describe the neutron-target interactions, it is important to look at the PKA spectrum. More relevantly, the weighted average recoil spectrum (WARS) should be considered, as it allows getting a reasonable description of the defect production process, weighing the primary recoil spectrum by the damage energy produced in each recoil event [15,19]. The median energy, $T_{1/2}$, which is the energy for which half of the recoils are generated in cascades with energies greater than $T_{1/2}$ (and half with energies lower), is frequently used as a sound parameter to describe a given particle-solid interaction process [19]. Even though both the WARS and the median energy do not provide a complete description of the recoil spectrum, they undeniably carry more information than the sole estimation of the number of displaced atoms. Besides, the choice of the ion nature and energy can be made so that the corresponding WARS matches that of the neutron spectrum inside a given reactor. This is precisely what we mean by "targeted irradiations" (see introduction).

In order to calculate the WARS corresponding to a neutron spectrum, we used the DART code (whose detailed description is given in [20]). Briefly, this code was created to estimate the PKA spectra produced in a polyatomic solid target. The neutron cross-sections are directly extracted from neutron libraries for each isotope (ENDF /B VI in the standard version), and in order to take into account resonances in the differential neutron isotope cross-section, a multi group approach was chosen to compute the PKA spectrum. The nuclear reactions are also considered (as the one taking place with Ni, see below). The DART code treats ion-solid interactions within the Binary Collision Approximation (BCA) framework [21, 22]. For nuclear stopping, the repulsive Thomas Fermi interatomic potential was used and for electronic stopping, the Ziegler formalism was implemented. The WARS is one of the direct output files of DART. An example is given in Fig.1 that displays, in particular, the spectrum corresponding to the Russian reactor IVV-2M, using $E_d(Ni)$ = 40 eV (note that this is a commonly used value, but in fact, the actual absolute value is meaningless as long as we use the same one for ion-target and neutron-target interaction simulations). For this reactor, the median energy, $T_{1/2}$, is 57.3 keV. It is worth mentioning that we included the nuclear reactions occurring between $^{58}$Ni and $^{59}$Ni isotopes and thermal neutrons, which is why we used the entire neutron spectrum of the reactor (this spectrum is shown in Appendix A).

To calculate the WARS of an ion-target experiment, we used both DART and SRIM codes. SRIM being an extremely well-known package of codes designed for computing any ion-target interaction, we will not present it here; details can be found in [23]. It is however important to mention that, if the WARS is a direct output of DART, it is not the case for SRIM. In order to obtain such a spectrum, one needs to process the (heavy) text file named "collisions.txt". We wrote Python scripts that allow taking this file as an input and getting the WARS as an output (see Appendix C).

In addition to the WARS associated to the reactor, Figure 1 presents three WARS corresponding to the interaction between 600 keV $^{58}$Ni$^+$ ions and a nickel target: one spectrum was obtained with the DART code while the two others come from SRIM using the two options "Quick calculations or Kinchin-Pease, KP" and "Full damage cascades, FC". There is almost no difference between these two later spectra (which is expected, given the definition of the WARS), whereas that derived from DART exhibits a significant discrepancy. We verified (not shown here) that this difference does not originate from different stopping powers between the two codes. In fact, this difference comes from the fact that DART only uses the initial incident ion energy to solve the analytical equation giving the WARS (which makes sense for neutrons), but SRIM follows the projectile all along its path and takes into account its slowing-down (which is sound for ions). Consequently, the whole spectrum derived from SRIM is shifted towards lower energies as compared to DART. In the following, depending on the cases, we will use one or the other code to determine the WARS for an ion-target interaction (and DART for all neutron-target interactions).

Another important parameter to consider when comparing ion and neutron irradiation experiments is the exposure "dose". There exist discrepancies between the number of displaced atoms (estimated within the BCA, like with the KP model for instance) and the actual defect density (experimentally measured or determined by simulation methods); this issue is well documented [16,18]. The quantification of actually produced defects requires a huge amount of experimental work and/or computational efforts [15-18,24-25]. This difficulty prevents one from using the defect density as a simple, common parameter to compare irradiation experiments. The fluence does not appear as a relevant parameter either, as it does not integrate the displacement cross-section nor does it describe the collision characteristics. We thus used the displacement per atoms (dpa) parameter (even though a new concept has recently emerged, the arcdpa [25], that we do not use here because it requires a lot of fundamental data to be used). For DART calculations, the neutron spectrum is known and the dpa production rate is an output value. Note that the estimation of the number of displaced atoms does not rely on the Kinchin-Pease formalism in DART [26], it is based on the Lindhard formalism [27]. Therefore, for SRIM, the calculation of the average dpa was derived from the 'collision events' distributions determined in the FC mode. We are aware that there exists a controversy on the 'correct' way to estimate the dpa values with SRIM [28-30]. However, in the present work, we only needed a damage parameter that could be used for both ion and neutron irradiation experiments, irrespective of its actual significance and potential agreement with actual defect concentrations. Consequently, in both cases, the dpa values were based on the number of displaced atoms, a quantity that we did specifically calculate ourselves for each (ion or neutron) irradiation condition.

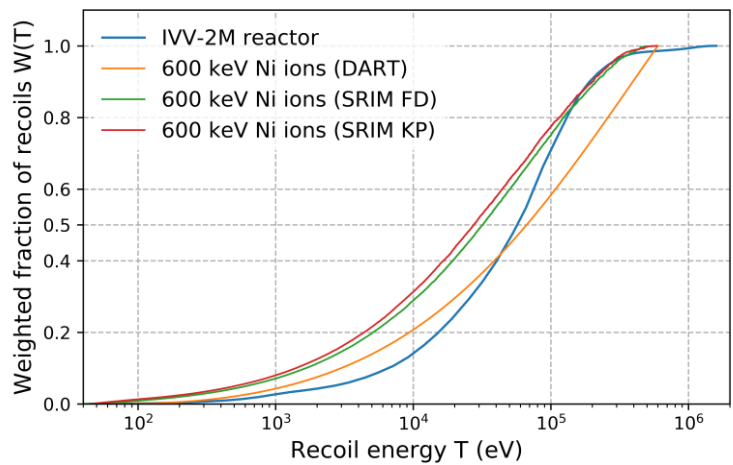

*Figure 1: Weighted average recoil spectra corresponding to the neutron spectrum of the IVV-2M reactor and to 600 keV $Ni^+$ ions. Calculations were performed with either DART [20] or SRIM [23] in "Full damage, FD" or "Kinchin-Pease, KP" modes, as indicated in the legend.*

## III. Results and discussion

### III. 1. Ni target in the IVV-2M REACTOR

In 2016 was published a paper that presented the lattice parameter change measured by neutron diffraction in Ni single crystals irradiated in the Russian IVV-2M reactor (see [31] and Fig.1 in Appendix B). We decided to try to reproduce these data performing ion irradiation experiments. In Fig.2 are presented the WARS corresponding to the nuclear reactor, along with those determined with DART for both 300 keV $Bi^{2+}$ and 600 keV $Ni^+$ ions, using $E_d(Ni)$ = 40 eV. Note that, as explained above, WARS for ion-target interactions should have been determined with SRIM instead of DART, but for this case of Ni irradiated samples, as it will be shown below, using one or the other would have had little effect on the experimental results (besides, a comparison between ion irradiation experiments with ions for which WARS have been calculated with both SRIM and DART is conducted in section III.3.).

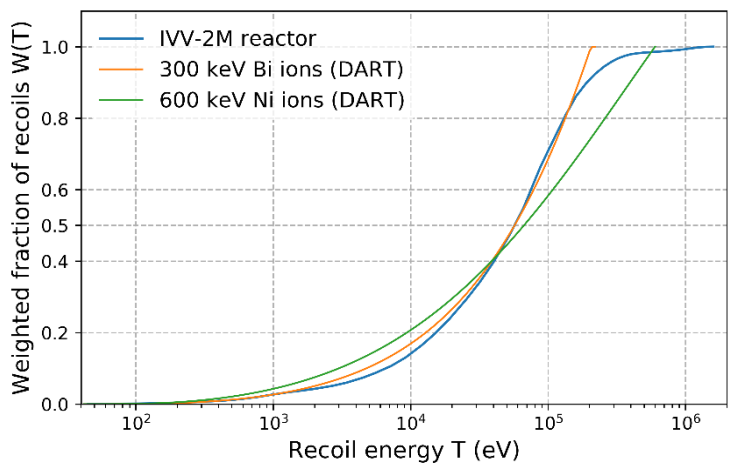

*Figure 2: Weighted average recoil spectra corresponding to the neutron spectrum of the IVV-2M reactor and to 600 keV $Ni^+$ and 300 keV $Bi^{2+}$ ions. Calculations were performed with DART [20].*

From Fig.2, it can be seen that the WARS of the 300 keV $Bi^{2+}$ ions is very close to that of the IVV-2M reactor, not only in terms of the median energy (~57 keV for both) but also with respect to the global shape. The WARS of the 600 keV $Ni^+$ ions appears different, particularly above $T_{1/2}$, but we carried out those Ni irradiations to increase the irradiated thickness (see Fig.3 for the dpa profiles). Table I summarizes the irradiation conditions. XRD results for Bi and Ni irradiations are presented in Fig.4a and Fig.4b, respectively. Clearly, no lattice parameter change is detectable for Bi irradiations, as no peak shift is observed (Fig. 4a). After significantly increasing the damage thickness with the Ni irradiations and also doubling the dpa level (with respect to the maximum dpa inside the reactor), this result still holds (Fig. 4b). Only some diffuse scattering due to the presence of defects is observed. The lattice parameter change measured after neutron irradiation was very weak (see Fig.2 of Appendix B and [31]), but given the resolution we have, we should have been able to detect such low strain levels, at least at the highest fluences and for the Ni irradiations, for which the damaged thickness is ~350 nm.

This statement is based on simulations of XRD curves using the RaDMaX-online program which is dedicated to dealing with irradiated materials containing strain depth profiles [32]. But it is not so surprising to not measure any strain, as we faced the same issue in another work on ion-irradiation in Ni single crystals [33]. This result nevertheless indicates that ion irradiation cannot allow straightforwardly emulating any neutron irradiation experiment.

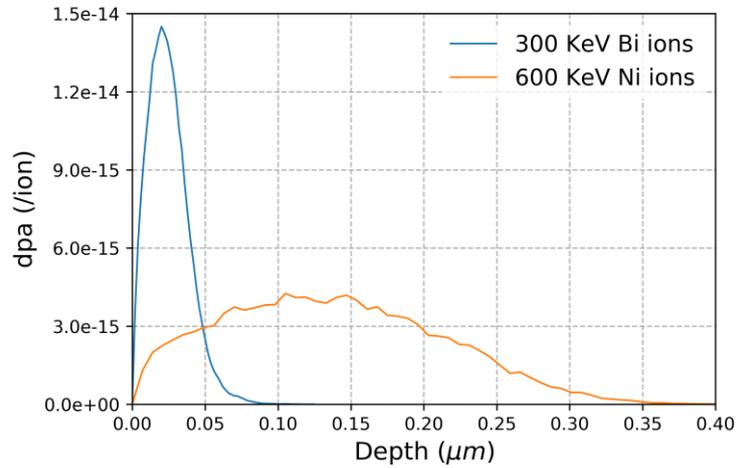

*Figure 3: Dpa depth distributions corresponding to the two conditions implemented for the irradiation of the Ni single crystals.*

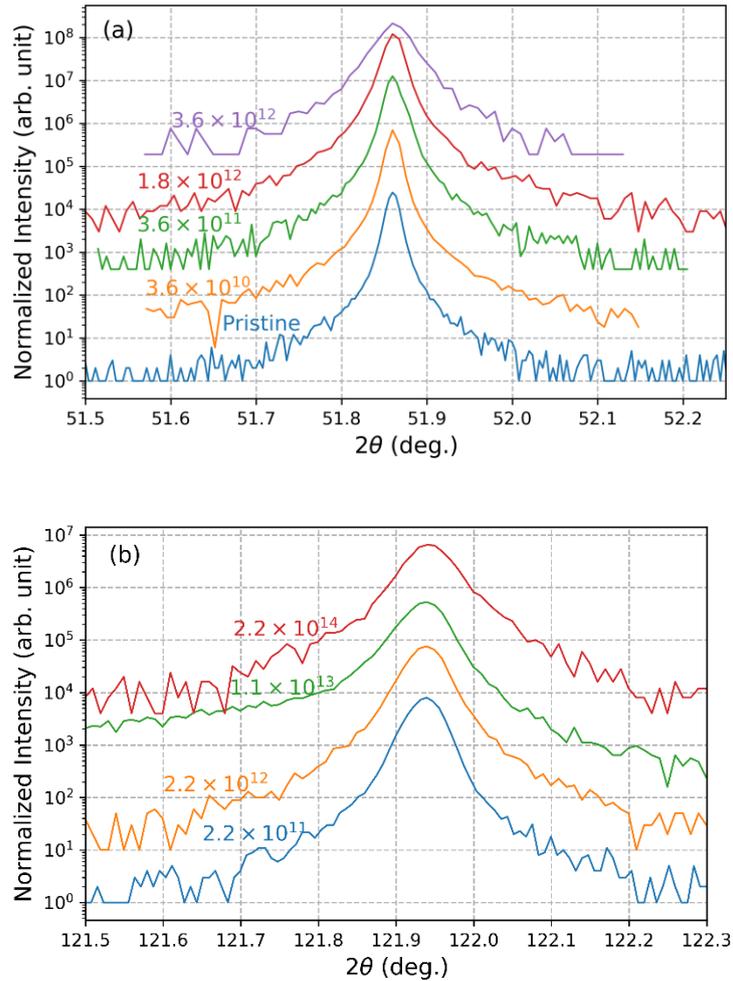

*Figure 4: XRD curves of Ni single crystals irradiated with a) 300 keV $Bi^{2+}$ ions and b) 600 keV $Ni^+$ ions at the indicated fluences. Curves are stacked for visualization purposes.*

### III.2. Cu target in the OAK RIDGE BULK SHIELDING REACTOR

In 1974, B. C. Larson reported a work on the lattice parameter changes in copper irradiated inside the Oak Ridge bulk shielding reactor, ORBSR (see [34] and Fig.2 in Appendix B). We decided to try to reproduce these data performing ion irradiation experiments. We were aware that the strain measured by Larson was very weak, but we expected to detect some strain at higher dpa level. In Fig.5 are presented the WARS, using $E_d(Cu) = 25$ eV, corresponding to the nuclear reactor, along with that determined with SRIM (not with DART this time) for 1.5 MeV $I^+$ ions. The two spectra display a similar shape and a very close median energy, around 75 keV. We also performed ion irradiations with 4 MeV $Au^{2+}$ ions; the corresponding WARS is shown in Fig.5. It does not match that of the reactor, but these ions have been chosen for two reasons: to reach a higher dpa level (see comment above) without having to increase too much the ion fluence, and in the meantime to extend the damage thickness (see Fig.6). The dpa levels and ion fluences are summarized in Table II. In addition, we carried out irradiations with iodine ions of different energies in order to have a larger, quasi-homogeneous disorder depth profile (Fig.6). For these irradiations of Cu samples, as for Ni crystals, no lattice parameter change has been measured, as evidenced by XRD results displayed in Figs.7 and 8. The only notable result can be obtained from Fig.8 that corresponds to data for the multi-energy ion-irradiation experiment. Indeed, there is a signal arising, at high ion fluence, on the high-angle side, which suggests either diffuse scattering coming from interstitial-type defects or negative elastic strain due to vacancy-type defects that would be dominant (over interstitials). It is not in the scope to further interpret these part of the data, but it is worth mentioning that they can be fitted (not shown here) assuming a 1 µm layer, that corresponds to the SRIM-predicted thickness, exhibiting a compressive -0.08 % strain and a high degree of disorder (static Debye-Waller factor of 0.25, instead of 1 for a perfect lattice).

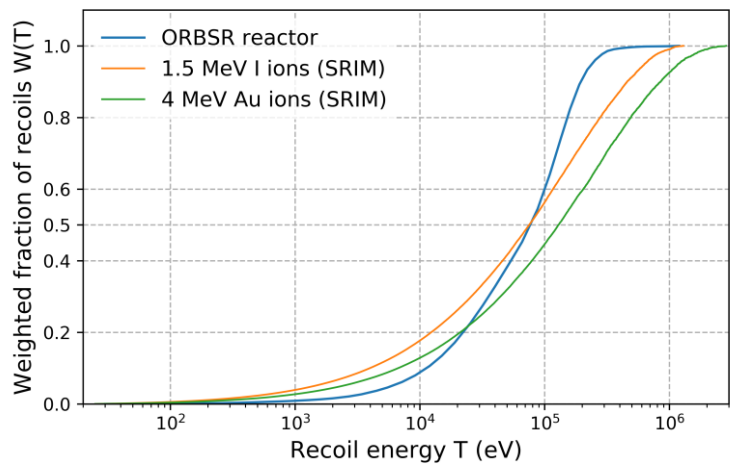

*Figure 5: Weighted average recoil spectra corresponding to the neutron spectrum of the ORBSR reactor and to 1.5 MeV I$^+$ and 4 MeV Au$^{2+}$ ions. Calculations were performed with SRIM [23].*

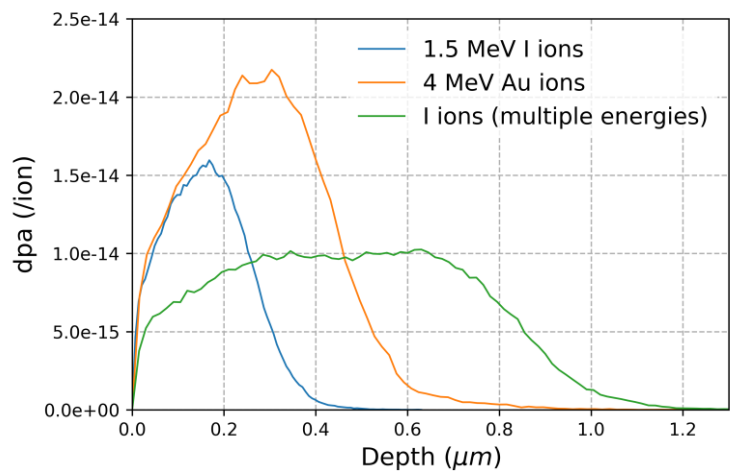

*Figure 6: Dpa depth distributions corresponding to the three conditions implemented for the irradiation of the Cu single crystals.*

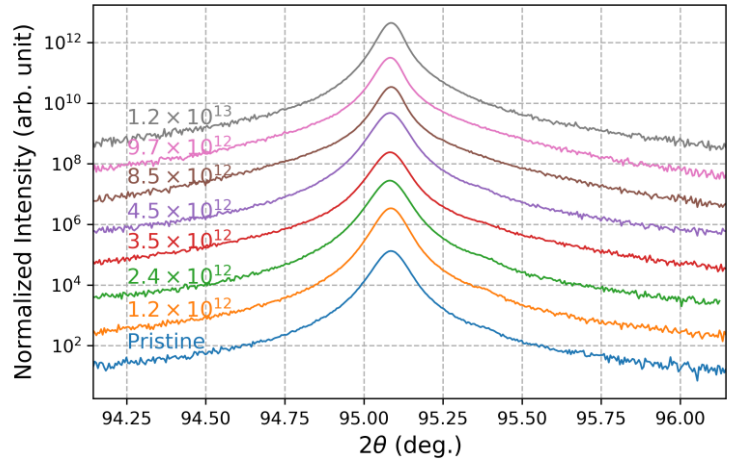

*Figure 7: XRD curves of Cu single crystals irradiated with 1.5 MeV I$^+$ ions at the indicated fluences. Curves are stacked for visualization purposes.*

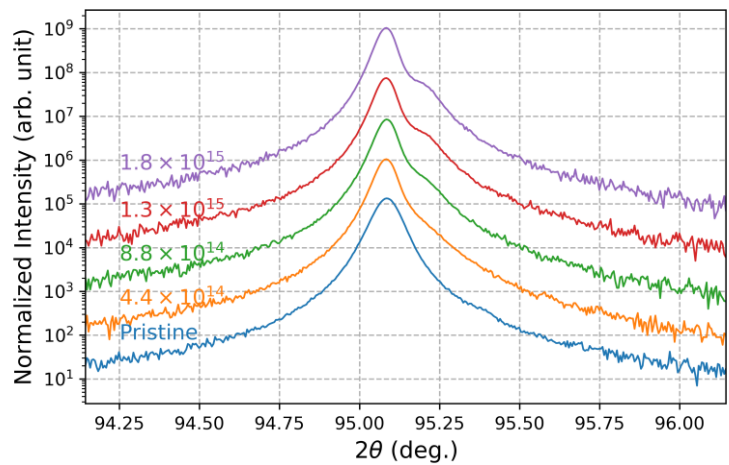

*Figure 8: XRD curves of Cu single crystals irradiated with I$^+$ ions of different energies (in order to have a flat disorder profile, see text) at the indicated total fluences. Curves are stacked for visualization purposes.*

This discrepancy in the measured elastic strain for both Ni and Cu crystals between neutron and ion irradiation experiments can be explained by a too small damaged volume generated during ion irradiation, particularly because the expected elastic strain is very weak. Therefore, even though the actual microstructure of the ion-irradiated crystals might be similar to that produced after neutron irradiation, different characterization techniques such as transmission electron microscopy must be used to proceed to a comparison. The monitoring of the lattice parameter, which has the advantage of being achievable without any sample preparation, is not appropriate in these cases and most likely in any situation where pure metals are involved (the situation can be different for alloys, as shown in [33]). On the contrary, ceramic materials exhibit much larger strain values when subjected to irradiation [35]. This is the case for instance for magnesium oxide, MgO.

**III.3. MgO target in the DIDO REACTOR**

In 1971 was published a work presenting the change in lattice parameter of MgO single crystals irradiated in the Harwell DIDO reactor [36]. We decided to try to reproduce these data performing ion irradiation experiments. In Fig.9 are presented the WARS corresponding to the nuclear reactor, along with those determined for 1 MeV Cl$^+$ ions with both DART and SRIM, and the one for 1.5 MeV I$^+$ ions (from SRIM); we used $E_d$(Mg) = $E_d$(O) = 60 eV. One can see that the WARS for 1 MeV Cl$^+$ ions determined with DART matches that of the DIDO reactor, but looking at the corresponding one determined with SRIM, there is a discrepancy (as explained previously). Therefore, we also performed 1.5 MeV I$^+$ irradiations (as for Cu) for which the WARS is closer to that of the reactor; in particular, the corresponding $T_{1/2}$ are 19 keV and 17.5 keV, respectively. To finish, we show in Fig. 9 the WARS for 1.2 MeV Au$^+$ ions (because we had a significant amount of data regarding MgO crystals irradiated with these ions), and (ii) the WARS for 1 MeV Ne$^+$ because we used corresponding data from the literature. The dpa depth distributions for Cl$^+$, I$^+$ and Au$^+$ ions are presented in Fig.10 and the dpa levels and ion fluences are summarized in Table III.

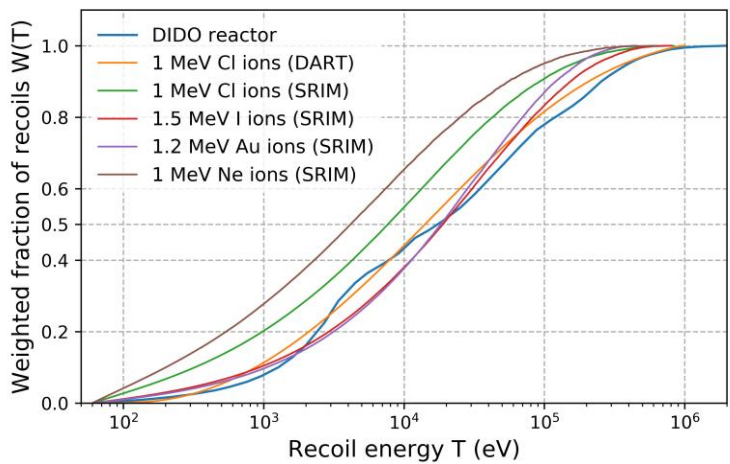

*Figure 9: Weighted average recoil spectra corresponding to the neutron spectrum of the DIDO reactor and to 1 MeV Cl$^+$, 1.5 MeV I$^+$, 1.2 MeV Au$^+$ and 1 MeV Ne$^+$ ions. Calculations were performed with either SRIM [23] or DART [20], see legend.*

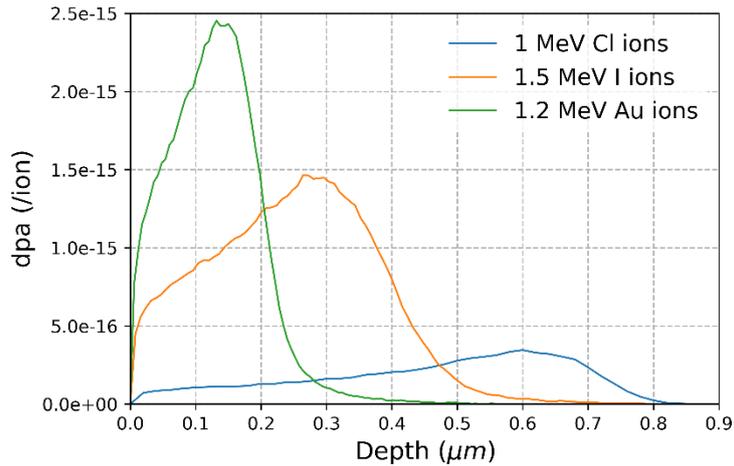

*Figure 10: Dpa depth distributions corresponding to the three conditions implemented for the irradiation of the MgO single crystals*

Figs.11 and 12 show XRD curves of MgO crystals irradiated with 1 MeV Cl$^+$ and 1.5 MeV I$^+$ ions, respectively; data for Au-irradiated samples were already presented in [24,37]. For this material, we do measure an elastic strain, as indicated by the appearance of a diffraction signal on the low-angle side of the curves [38]. This strain is tensile, in agreement with the lattice expansion observed after irradiation in the DIDO reactor [36]. The strain magnitude as a function of dpa level is plotted in Fig. 13 for all irradiation conditions, including the neutron irradiation. Note that the strain values for ion-irradiated crystals cannot be determined directly from the raw XRD data. One must use a mechanical model to take into account the fact that the ion-irradiated layers, contrary to bulk crystals in a neutron reactor, are free to swell only along the surface normal direction because in their plane, there exists a constraint from the thick, rigid unirradiated part underneath them. The validity of this method has been demonstrated several times in e.g. [38-39].

Clearly, the strain kinetics in the case of iodine irradiation does not match the lattice parameter change determined for neutron-irradiated samples (see [36] and Fig.13). Yet, 1.5 MeV I$^+$ ions had a very similar WARS as compared to that experienced by the MgO crystals in the DIDO reactor. The strain kinetics for both 1 MeV Cl$^+$ (at RT) and 1.2 MeV Au$^+$ (at 80 K) ion irradiations do not agree either with the one of the neutron-irradiated samples. In particular, the strain levels for ion-irradiated crystals are too large. The three curves (for I, Cl and Au) exhibit a very similar trend and strain values are very close. This resemblance could be due to similar WARS. In Fig.13 is also plotted the strain as a function of dpa for MgO samples irradiated with 1.2 MeV Au$^+$ ions at 1073 K. The corresponding strain kinetics matches that determined for neutron irradiations. It has been shown that increasing the temperature during irradiation leads, in MgO, to a decrease in the defect production rate (the defects produced in collision cascades are more annealed out because of the thermal energy) [37]. Hence, at 1073 K, the strain levels are lower than at RT, and the values are in agreement with those obtained after irradiation in the DIDO reactor, even though these irradiations occurred at 300 K. It is known, particularly for metals, that the difference in dpa rate between neutron and ion irradiation experiments can be accounted for by an increase in temperature during ion irradiation, so that the actual net defect fluxes are better reproduced. In the present case, it appears that a dynamic annealing must take place (to reach a lower defect density), but this is achieved by a significantly large temperature shift of nearly 3.5 times the temperature inside the reactor. However, such an important shift in temperature has already been reported to account for a large difference in damage rates - on the order of $10^{12}$ - between experiments and calculations in irradiated (ceramic oxide) pyrochlores [40]. The difference in dose rate between neutron and Au irradiations is on the order of $10^9$, so a large temperature shift for a material of the same class (ceramic oxide) appears plausible.

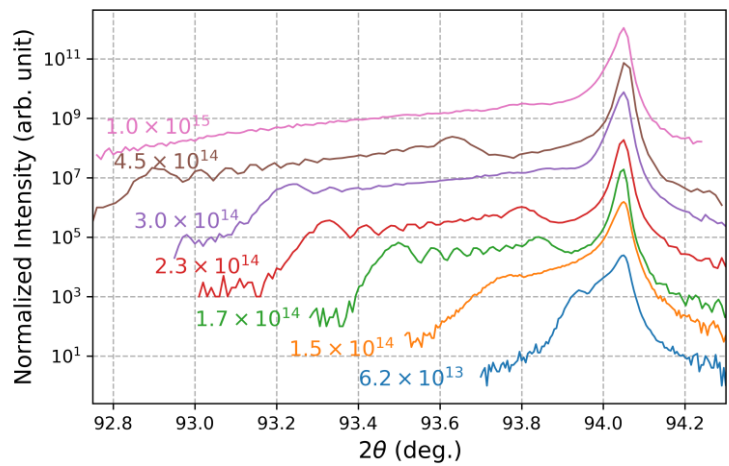

Figure 11: XRD curves of MgO single crystals irradiated with 1 MeV Cl$^+$ ions at the indicated fluences. Curves are stacked for visualization purposes.

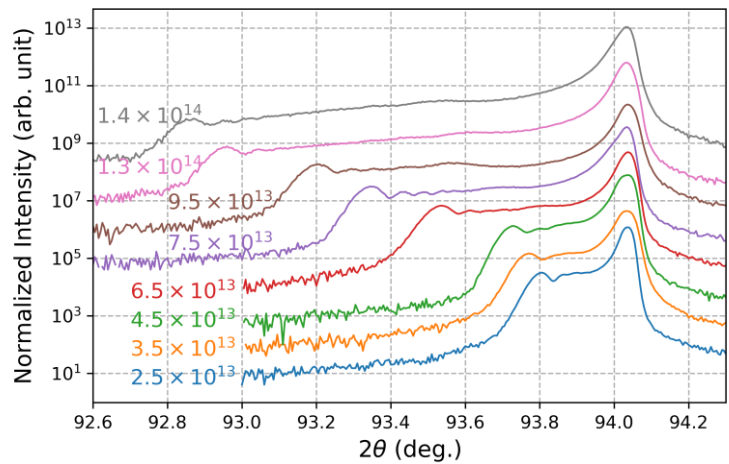

Figure 12: XRD curves of MgO single crystals irradiated with 1.5 MeV I$^+$ ions at the indicated fluences. Curves are stacked for visualization purposes.

For the purpose of comparison between dpa values estimated from "Quick-Calculation" and 'Full Cascades' modes, we plotted in Fig. 14 the same strain kinetics as in Fig.13, but with dpa levels calculated using the quick mode. Except that KP-dpa values are about 20 % lower than the FC-dpa ones, there is no substantial change and the agreement between the high-temperature Au irradiations and the neutron irradiations remains extremely close. In Fig.14, we also plot data that we adapted from [41] regarding 1 MeV Ne$^+$ irradiation of MgO single crystals. We recalculated the average KP-dpa values. We can observe that the strain kinetics is almost identical to the set of the three curves corresponding to chlorine, iodine and gold (at 80 K) irradiations. Yet, the WARS of Ne$^+$ ions is very different to the three others, notably the median energy which is much smaller (4 keV *vs* 17 keV). This result suggests that, to simply monitor the lattice parameter change, in a material like MgO and most likely in many other ceramic materials that do not amorphize, the most important irradiation parameter is the effective defect creation rate. Nonetheless, the actual microstructure might not be that obtained after neutron irradiation if the cascade features are too different, hence, taking care of the WARS should be considered.

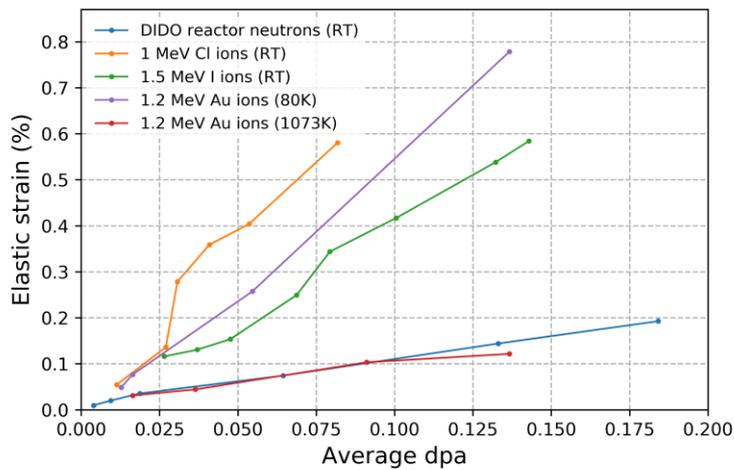

*Figure 13: Strain values measured in irradiated MgO single crystals under the conditions indicated in the legend as a function of the dpa level determined with SRIM using the FC mode.*

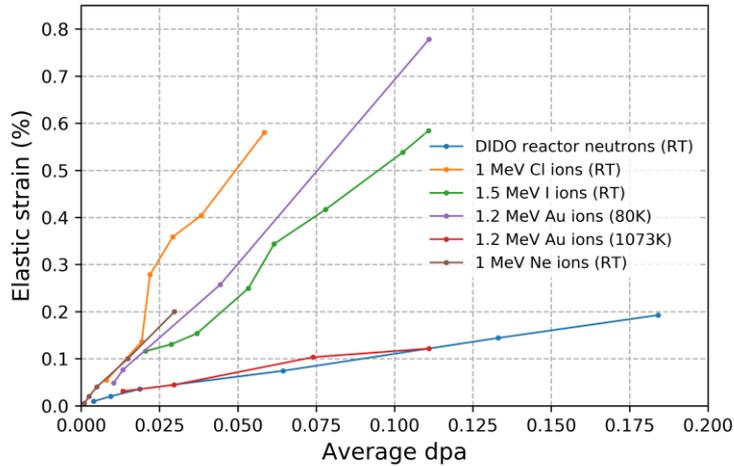

*Figure 14: Strain values measured in irradiated MgO single crystals under the conditions indicated in the legend as a function of the dpa level determined with SRIM using the KP mode.*

**Discussion and conclusion**

In this work, we performed ion irradiation experiments specifically designed to mimic neutron irradiations. We selected three materials, namely Ni, Cu and MgO single crystals, for which there exists, in the literature, data of lattice parameter change (elastic strain) after irradiation in a nuclear reactor. The choice of the ion nature and energy was dictated by the aim of having similar weighted average recoil spectra for ion and neutron irradiation experiments. We then performed HR-XRD measurements to monitor the strain build-up. We found that for Ni and Cu (and presumably any pure metal), ion irradiation cannot allow, as straightforwardly as here tempted, reproducing experimental results like elastic strain developing upon neutron irradiation. On the contrary, for MgO, ion irradiation can lead to similar results as those found after irradiation in a reactor, providing that the (ion) irradiation temperature is significantly increased in order to account for the large discrepancy in the damage creation rates. Therefore, choosing an *ad hoc* WARS is not sufficient to ensure to perfectly mimic neutrons with ions.

As a final remark, we would suggest the following procedure in order to use ion irradiation as a proxy for neutron irradiation:

(1) Find an ion (nature and energy) for which the WARS is close to that associated with the neutron spectrum in the reactor
(2) Monitor a simple quantity, as the elastic strain or any damage level parameter, to get a disordering kinetics
(3) Adapt ion fluence and/or temperature to match the neutron data
(4) Solely then, use techniques more complex to implement, such as electron microscopy or atomic probe tomography.

Steps (1) to (3) should allow saving time by preventing from implementing destructive, not straightforward techniques, particularly if they are combined with predictions from computational techniques such as phase field modeling. Step (4) will allow to ensure that the actual microstructure is similar to that obtained after neutron irradiation

**Acknowledgments**

XJ and AD would like to thank Laurence Luneville and David Simeone, CEA Paris-Saclay, for fruitful scientific discussions about the comparison between SRIM and DART codes.

**References**

[1] G. Was, J. Mater. Research 30 (2015) 1158.

[2] S. J. Zinkle, L. L. Snead, Scripta Mater. 143 (2018) 154.

[3] M. Roldán, P. Galán, F. J. Sánchez, I. García-Cortés, D. Jiménez-Rey, Pilar Fernández, IntechOpen, Online First (2019), DOI: 10.5772/intechopen.87054.

[4] L. K. Mansur, J. Nucl. Mater. 206 (1993) 306.

[5] C. Abromeit, J. Nucl. Mater. 216 (1994) 78.

[6] O. V. Ogorodnikova, V. Gann, J. Nucl. Mater. 460 (2015) 60.

[7] N. Galy, N. Toulhoat, N. Moncoffre, Y. Pipon, N. Bérerd, M. R. Ammar, P.Simon, D. Deldicque, P. Sainsot, J. Nucl. Mater. 502 (2018) 20.

[8] R. W. Harrison, Vacuum 160 (2019) 355.

[9] L. L. Snead, Y. Katoh, T. Koyanagi, K. T. Eliot, D. Specht, J. Nucl. Mater. 471 (2016) 92.

[10] S. Jublot-Leclerc, X. Li, L. Legras, M.-L. Lescoat, F. Fortuna, A. Gentils, J. Nucl. Mater. 480 (2016) 436.

[11] Z. Jiao, J. Michalicka, G. S. Was, J. Nucl. Mater. 501 (2018) 312.

[12] M. Hernández Mayorala, F. Bergner, C. Heintze, V. Kuksenko, C. Pareige, Ph. Pareige, L. Malerba, NEA-NSC-WPFC-DOC--2015-9 (RN:46040920).

[13] Ce Zheng, M. A. Auger, M. P. Moody, D. Kaoumi, J. Nucl. Mater. 491 (2017) 162.

[14] A. Debelle, J. Channagiri, L. Thomé, B. Décamps, A. Boulle, S. Moll, F. Garrido, M. Behar, J. Jagielski, J. Appl. Phys 115 (2014) 183504.

[15] R. S. Averback, R. Benedek, K. L. Merkle, Phys. Rev. B 18 (1978) 4156.

[16] J-P Crocombette, L. Van Brutzel, D. Simeone, L. Luneville, J. Nucl. Mater. 474 (2016) 134.

[17] C. J. Ortiz, Comp. Mater. Science 154 (2018) 325.

[18] K. Nordlund, A. E. Sand, F. Granberg, S. J. Zinkle, R. Stoller, R. S. Averback, T. Suzudo, L. Malerba, F. Banhart, W. J. Weber, F. Willaime, S. Dudarev, D. Simeone, J. Nucl. Mater. 512 (2018) 450.

[19] R. S. Averback, J. Nucl. Mater. 216, 49 (1994).

[20] L. Lunéville, D. Simeone, C. Jouanne, J. Nucl. Mater. 353 (2006) 89.

[21] M. T. Robinson, Rad. Effects and Defect in Solids Null (1994) 3.

[22] W. Eckstein, The Binary Collision Model in "Computer Simulation of Ion-Solid Interactions", p.4, Springer, Berlin, Heidelberg, 1991.

[23] J. F. Ziegler, J. P. Biersack, U. Littmark, The Stopping and Range of Ions in Solids, Pergamon, New York, 1985. SRIM program can be downloaded at: www.srim.org.

[24] D. Bachiller-Perea, A. Debelle, L. Thomé, M. Behar, J. Nucl. Mater. 478 (2016) 268.

[25] K. Nordlund, S. J. Zinkle, A. E. Sand, F. Granberg, R. S. Averback, R. Stoller, T. Suzudo, L. Malerba, F. Banhart, W. J. Weber, F. Willaime, S. L. Dudarev, D. Simeone, Nature Comm. 9 (2018) 1084.

[26] L. Luneville, D. Simeone, D. Gosset, Nucl. Instr. and Methods B 250 (2006) 71.

[27] J. Lindhard, Phys. Rev. B 14 (1961) 1.

[28] R. E. Stoller, M. Toloczko, G. S. Was, Nucl. Instr. and Methods B 310 (2013) 75.

[29] J-P Crocombette, C. Van Wambeke, EPJ Nuclear Sci. Technol. 5 (2019) 7.

[30] W. J. Weber, Y. Zhang, Current Opinion Solid State and Materials Science 23 (2019) 100757.

[31] V. I. Voronin, I. F. Berger, N. V. Proskurnina, B. N. Goschitskii, The Physics of Metals and Metallography 117 (2016) 348.

[32] A. Boulle, V. Mergnac, J. Appl. Crystal., in press.

[33] N. Sellami, A. Debelle, M. W. Ullah, H. M. Christen, J. K. Keum, H. Bei, H. Xue, W. J. Weber, Y. Zhang, Current Opinion in Solid State and Materials Science 23 (2019) 107.

[34] B. C. Larson, J. Appl. Phys. 45 (1974) 514.

[35] A. Debelle, J.-P. Crocombette, A. Boulle, A. Chartier, Th. Jourdan, S. Pellegrino, D. Bachiller-Perea, D. Carpentier, J. Channagiri, T.-H. Nguyen, F. Garrido, L. Thomé, Phys. Rev. Mater. 2 (2018) 013604.

[36] B. Henderson, D. H. Bowens, J. Phys. C: Solid St. Phys. 4 (1971) 1487.

[37] A. Debelle, J.-P. Crocombette, A. Boulle, E. Martinez, B. P. Uberuaga, D. Bachiller-Perea, Y. Haddad, F. Garrido, L. Thomé, M. Béhar, Phys. Rev. Mater. 2 (2018) 083605.

[38] A. Debelle, A. Declémy, Nucl. Instr. and Methods B 268, (2010) 1460.

[39] A. Debelle, A. Boulle, F. Rakotovao, J. Moeyaert, C. Bachelet, F. Garrido, L. Thomé, J. Phys. D: Appl. Phys. 46, (2013) 045309.

[40] J-P Crocombette, A. Chartier, W. J. Weber, Appl. Phys. Lett. 88 (2006) 051912.

[41] A. I. Van Sambeek, R. S. Averback, Mat. Res. Soc. Symp. Proc. Vol. 396 (1996) 137.

Table I: Detailed characteristics of neutron and ion irradiations

| IVV-2M neutron fluence (cm$^{-2}$) | dpa | 300 keV Bi ions (cm$^{-2}$) | dpa | 600 keV Ni ions (cm$^{-2}$) | dpa |
|---|---|---|---|---|---|
| 10$^{18}$ | 4.3x10$^{-4}$ | 3.6x10$^{10}$ | 3.9x10$^{-4}$ | 2.15x10$^{11}$ | 7.1x10$^{-4}$ |
| 10$^{19}$ | 4.3x10$^{-3}$ | 3.6x10$^{11}$ | 3.9x10$^{-3}$ | 2.15x10$^{12}$ | 7.1x10$^{-3}$ |
| 5x10$^{19}$ | 2.15x10$^{-2}$ | 1.8x10$^{12}$ | 1.95x10$^{-4}$ | 1.1x10$^{13}$ | 3.55x10$^{-2}$ |
| 10$^{20}$ | 4.3x10$^{-2}$ | 3.6x10$^{12}$ | 3.9x10$^{-2}$ | 2.15x10$^{13}$ | 7.1x10$^{-2}$ |

| ORBSR neutron fluence (cm$^{-2}$) | dpa | 1.5 MeV I ions (cm$^{-2}$) | dpa | 4 MeV Au ions (cm$^{-2}$) | dpa | Mixed I ions (cm$^{-2}$) | dpa |
|---|---|---|---|---|---|---|---|
| 3.6x10$^{18}$ | 1.45x10$^{-2}$ | 1.2x10$^{12}$ | 1.4x10$^{-2}$ | 3.45x10$^{12}$ | 5.6x10$^{-2}$ | 4.4x10$^{14}$ | 3.85 |
| 7.2x10$^{18}$ | 2.9x10$^{-2}$ | 2.35x10$^{12}$ | 2.9x10$^{-2}$ | 5.55x10$^{12}$ | 9x10$^{-2}$ | 8.8x10$^{14}$ | 7.65 |
| 1.1x10$^{19}$ | 4.4x10$^{-2}$ | 3.55x10$^{12}$ | 4.4x10$^{-2}$ | 7.35x10$^{12}$ | 1.2x10$^{-1}$ | 1.3x10$^{15}$ | 11.5 |
|  |  | 8.45x10$^{12}$ | 1.05x10$^{-1}$ | 9.2x10$^{12}$ | 1.5x10$^{-1}$ | 1.75x10$^{15}$ | 15.5 |
|  |  | 1.2x10$^{13}$ | 1.5x10$^{-1}$ |  |  |  |  |

| DIDO neutron fluence (cm$^{-2}$) | dpa | 1 MeV Cl ions (cm$^{-2}$) | dpa | 1.5 MeV I ions (cm$^{-2}$) | dpa |
|---|---|---|---|---|---|
| 1.7x10$^{18}$ | 3.95x10$^{-3}$ | 6.25x10$^{13}$ | 1.15x10$^{-2}$ | 2.5x10$^{13}$ | 2.65x10$^{-2}$ |
| 4.1x10$^{18}$ | 9.45x10$^{-3}$ | 1.5x10$^{14}$ | 2.7x10$^{-2}$ | 3.5x10$^{13}$ | 3.7x10$^{-2}$ |
| 8.1x10$^{18}$ | 1.85x10$^{-2}$ | 1.7x10$^{14}$ | 3.05x10$^{-2}$ | 4.5x10$^{13}$ | 4.75x10$^{-2}$ |
| 2.8x10$^{19}$ | 6.45x10$^{-2}$ | 2.25x10$^{14}$ | 4.1x10$^{-2}$ | 6.5x10$^{13}$ | 6.9x10$^{-2}$ |
| 5.8x10$^{19}$ | 1.35x10$^{-1}$ | 2.95x10$^{14}$ | 5.35x10$^{-2}$ | 7.5x10$^{13}$ | 7.95x10$^{-2}$ |
| 8x10$^{19}$ | 1.85 | 4.55x10$^{14}$ | 8.2x10$^{-2}$ | 9.5x10$^{13}$ | 10$^{-1}$ |
|  |  |  |  | 1.25x10$^{14}$ | 1.3x10$^{-1}$ |
|  |  |  |  | 1.35x10$^{14}$ | 1.45x10$^{-1}$ |

**Appendix A: Neutron spectra in the IVV-2M, ORBSR and DIDO nuclear reactors**

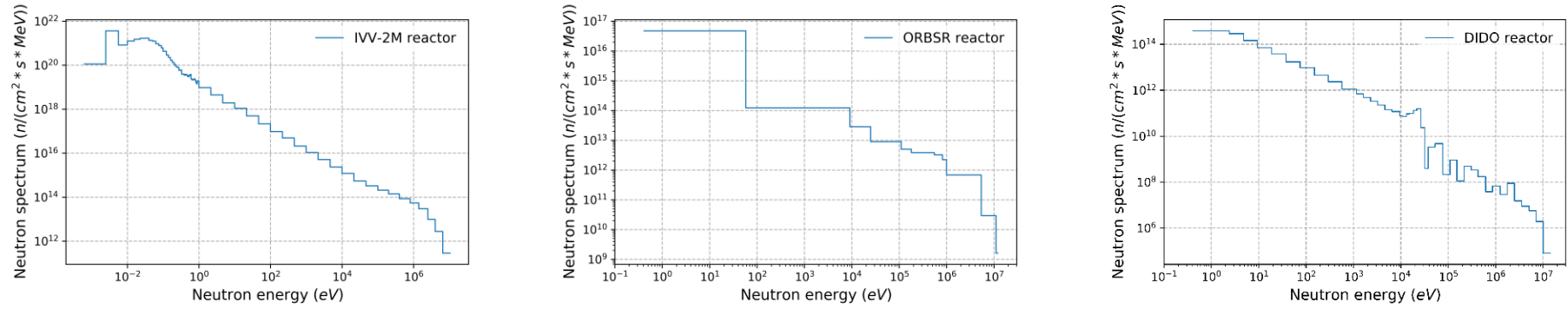

*Fig.1: Neutron spectrum for the three studied nuclear reactors*

**Appendix B: Lattice parameter change (strain) in Ni and Cu irradiated in the nuclear reactors**

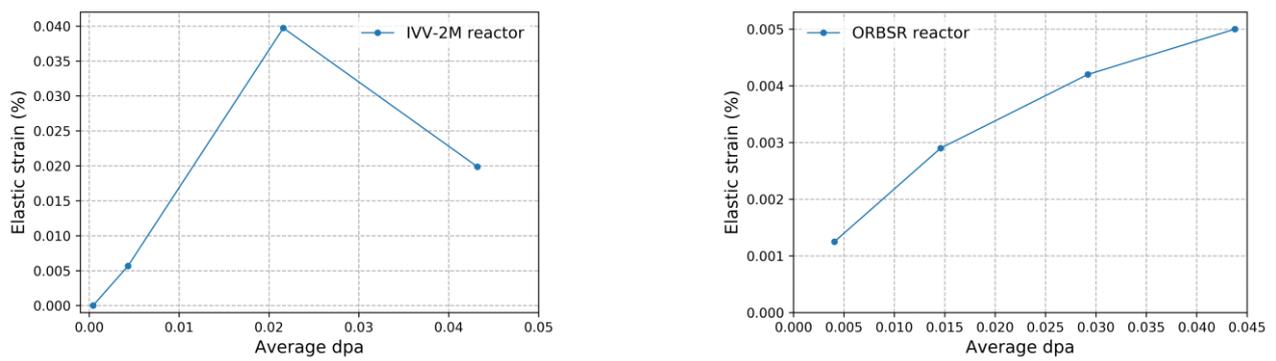

*Fig.2: Lattice parameter change (strain) as a function of the dpa level for Ni single crystals irradiated in the IVV-2M reactor (left) and Cu single crystals irradiated in the ORBSR reactor (right).*

**Appendix C: Python script to generate WARS from SRIM data**

In this Appendix is given the Python script (Numpy library) we wrote to use the SRIM "collisions.txt" file to generate a weighted average recoil spectrum.